# Focus on conceptual ideas in quantum mechanics for teacher training

**J K Freericks[1]**

[1]Department of Physics, Georgetown University, 37th and O Sts., NW, Washington, DC 20057
Email: james.freericks@georgetown.edu

**Abstract**

In this work, we describe strategies and provide case-study activities that can be used to examine the properties of superposition, entanglement, tagging, complementarity, and measurement in quantum curricula geared for teacher training. Having a solid foundation in these conceptual ideas is critical for educators who will be adopting quantum ideas within the classroom. Yet they are some of the most difficult concepts to master. We show how one can systematically develop these conceptual foundations with thought experiments on light and with thought experiments that employ the Stern-Gerlach experiment. We emphasize the importance of computer animations in aiding the instruction on these concepts.

## Introduction

The advent of the second quantum revolution—where individual quanta can be created, detected, and controlled—has the potential to transform the workplace to one that is quantum-enabled and quantum-enhanced. To prepare for this second revolution, we need a quantum-aware workforce that will be able to understand and work with quantum ideas. The teaching of quantum mechanics needs to move further down the curriculum and have placement in the primary and secondary education tracks. Key to making this work is the need to train educators, who often are quantum illiterate and possibly even afraid of quantum, in the ideas of quantum mechanics. In this paper, we argue that the most important element of teacher training is to develop a conceptual sense of what quantum mechanics is, and why it behaves how it does. The development of a conceptual knowledge base, before moving into more formal developments, is one of the most important conclusions to come from the physics education research field, as the physics education group group at the University of Washington and led by Lillian McDermott showed us with their *Tutorials in Introductory Physics* [1], which focused on developing conceptual ideas in the first-year physics course. It is our belief that quantum mechanics is no exception to this rule. The conceptual ideas must be developed *first* for teachers to understand the material well, so they can, in turn, teach their own students. In this work, we describe five conceptual ideas that we believe should be mastered and we describe in detail some methodologies that can be employed to teach these ideas to teachers along with learning goals to ensure their mastery. Successful completion of such a program will enable critical thinking within a quantum mindset. This is essential for educators to be able to teach the material to their students. It does not take years to master such material, but it cannot be done in a few day "bootcamp" either. It requires a dedicated period of time, at least 40 hours of work spread over four to six weeks to achieve this goal. But it requires no calculus, just an algebra and trigonometry background.

This paper is not a conventional physics education research article, it is more a call to action and a discussion of the key requirements and needs for educating teachers to become quantum literate. While education research has not been performed on these ideas yet, they have been used with many tens of thousands of students, and comments from the students have been employed to modify and improve the materials over the past five years. So, the work is both field tested and evidence informed, but not evidence based.

**The five conceptual ideas and common misconceptions**

The five critical ideas for conceptual understanding of quantum mechanics are superposition, entanglement, complementarity, tagging, and measurement. While most of these ideas might be familiar, we carefully define each of them. Quantum superposition occurs when a state vector is composed of more than one component, when expressed in a basis relevant for the current quantum system being studied. It is not a mathematically well-defined concept as an +$x$ oriented spin state is not in a superposition in an $x$-basis, but is in a $z$-basis. Nevertheless, by using an appropriate basis for a given situation, the concept is reasonably well defined. Entanglement is a special form of superposition, where a quantum state is described by more than one degree of freedom, and the quantum state vector cannot be factorized; it is often described as a state where the whole possesses correlations, not seen by any of its parts. Complementarity is described in terms of the concept of whether or not we have "which-way" information. In other words, is the system described by a simple superposition, or is it described by an entangled superposition, where the additional degree of freedom is correlated with the first degree of freedom and thereby provides which-way information? Tagging is the reversible creation of an entangled state that provides which-way information. Finally, measurement is an irreversible process, which can be destructive or not, and which can be used for detection or for state preparation. We now describe each concept more fully and explain why they are important and often misunderstood. We will also provide examples that can be employed to illustrate subtle points. In this work, our philosophical stance is that the wavefunction is a calculational tool, not a real dynamic quantity that "is" the quanta. We use the Copenhagen interpretation, within this philosophical mindset, for analyzing projective measurements. But we will also discuss counting measurements which do not fit that paradigm.

The hardest aspect of understanding superposition is the notion that a quantity may be indeterminant until measured. There are many ways to illustrate this behavior. Two particularly good ones are the Stern-Gerlach analyzer loop, discussed by Feynman in Volume III of the *Feynman Lectures in Physics* [2] and by Styer in *The Strange World of Quantum Mechanics* [3]. A similar device that illustrates this concept is Terry Rudolph's box that performs a Hadmard superposition in *Q is for Quantum* [4]. A third is the classic two-slit experiment (of course there are many others as well). We will first focus on the analyzer loop, as originally introduced by Wigner [5]. We will also discuss the two-slit experiment, especially as treated by Feynman in *QED: The strange theory of light and matter* [6].

An analyzer loop has an input port (on the left) and an output port (on the right, see Fig. 1). Inside the analyzer loop, the atoms are separated along two paths, which act like a Stern-Gerlach analyzer, correlating (or, if you like, entangling) the spatial position with the projection of the spin on the axis of the analyzer loop. The second part of the loop then rejoins the two paths back to one, erasing the separation and removing the entanglement. The beauty of this device is we can attach different "gates" at the maximal separation point. These ideas are developed with a minimum of math in Styer's book [3], and in an on-line class (that has had over 45,000 enrollees) [7,8]; javascript video tutorials [9] are available as well.

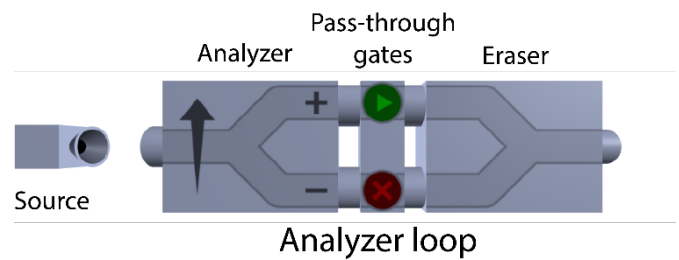

**Figure 1.** Schematic of an analyzer loop. It has one input and one output port. The source on the left, inputs atoms to the analyzer loop, entering the first Stern-Gerlach analyzer. The atoms separate and travel in superposition in each arm, and then rejoin via the eraser attachment. At the point of full separation, one can also attach different devices. Here, two pass-through gates are attached, one with a closed gate (bottom, red) and one with an open gate (top, green).

Figure 1 shows an analyzer loop oriented in the vertical $z$ direction, with an open pass-through gate on the positive projection arm and a closed pass-through gate on the negative projection arm. We denote spin states in Dirac notation as $|\uparrow\rangle_z$ for the positive projection on the z-axis and $|\downarrow\rangle_z$ for the negative projection on the z-axis. Similar notation is used for other axes. Suppose we send in an atom in the state $|\uparrow\rangle_x$ from the source. As it moves along the arms, it travels in superposition, given by $\frac{1}{\sqrt{2}}(|\uparrow\rangle_z + |\downarrow\rangle_z)$, since we use the eigenstates along the arms to represent that atom when inside the analyzer loop; one could also describe this in terms of an entangled state, with the spatial positions of the arms as the additional degree of freedom ($\frac{1}{\sqrt{2}}(|\uparrow\rangle \otimes |\text{top}\rangle + |\downarrow\rangle \otimes |\text{bottom}\rangle)$), but that is not required for most analyses. If both pass-through gates are open, the atom passes through and emerges in the $|\uparrow\rangle_x$ state. *The analyzer loop does nothing.* However, if one of the pass-through gates is closed, as in the picture, the $|\downarrow\rangle_z$ component is removed (which can be formally illustrated by the projection operator $|\uparrow\rangle_z {}_z\langle\uparrow|$ acting on the state), and we have the state $\frac{1}{\sqrt{2}}|\uparrow\rangle_z$ emerge from the analyzer loop; the reduction in amplitude of the state, represents the fact that only half of the atoms input into the analyzer loop emerge from the output (recall $_z\langle\uparrow|\uparrow\rangle_x = \frac{1}{\sqrt{2}}$)—in this case the state changes upon exit. This is an irreversible change of the state, which we associate with a measurement. The analyzer loop, and experiments constructed from it, create a powerful system that can be used to describe superposition. It does not require using Dirac notation, but using it can help in training teachers about how quantum mechanics works. Styer's text shows how to analyze situations using just high-school-level algebra and trigonometry [3].

One critically important concept is that when the atom travels in superposition, we cannot say whether it is on one arm or the other. We may want to say colloquially that it "travels on both arms," but it is much better to say that *precisely what it is doing is indeterminate*. There is no definite arm it moves on; it is not one arm or the other. In quantum mechanics we have another choice—*we do not know which arm it moves on*, This concept is quite difficult for students and teachers to grasp. It takes examples and time to learn. But mastering it is all that is needed to be able to work with the concept of superposition.

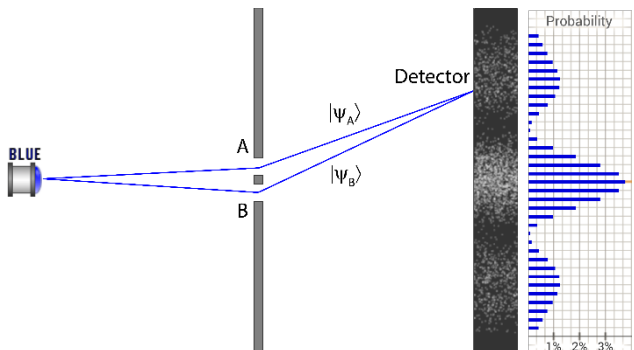

**Figure 2.** Schematic of a two-slit experiment. The blue source of light is on the left, the mask with two slits in the center, and the screen on the right (showing discrete photon "hits"). We display the two alternative ways to go from the source to the detector, and also illustrate the Dirac notation for the amplitudes for each way. The probability distribution is shown in blue on the far right and is given by $|\langle x|\psi_A\rangle + \langle x|\psi_B\rangle|^2$.

Another useful system is the two-slit experiment, which is depicted in Figure 2. Here, we have a monochromatic, single-photon source on the left, a screen with two narrow slits in it (labeled A and B) and a detector to the right. There are two alternative ways to go from the source to the detector, so the probability amplitude at the detector is given by $|\psi_A\rangle + |\psi_B\rangle$. For the detector located at position $x$, we take the inner product with the bra $\langle x|$ to determine the probability amplitude at the detector; normalization requires one to integrate the probability density $|\langle x|\psi_A\rangle + \langle x|\psi_B\rangle|^2$ over $x$ and introduce a normalization constant that makes the total integral equal to 1. Feynman's QED book [6] shows how to analyze these experiments using arrows that rotate according to the time it takes the photon to travel along the path, with a rotation rate given by the color of the photon. Calculus is not needed to analyze these systems; algebra and trigonometry suffice; the normalization can be handled discretely, by looking at the total number of arrows used in the description.

Both Stern-Gerlach and two-slit experiments are difficult to examine, but a thorough grounding in their behavior, as examined by Styer [3] and Feynman [6], allows one to discuss much complex quantum phenomena. These ideas have been incorporated into a 4-6 week long course on edX called *Quantum Mechanics for Everyone* [7]. When properly used, these materials can provide the background needed for a teacher to develop a conceptual understanding of quantum mechanics. But, it cannot be done too quickly. Quantum concepts are counterintuitive and hard. They take time to sink in. The online course employs a number of computer-based animations and tutorials, which help students grasp complex ideas and explore the systems they are working with.

One of the nice aspects of using these systems is one can easily explain the concept of complementarity with them. Complementarity is the principle that asks do we have which-way information, or not. If we do, then interference effects disappear. If we do not, then the different "paths" can interfere. One can illustrate these effects in either platform. A description for how to do this in the Stern-Gerlach apparatus is given in Courtney *et al.* [10] and won't be discussed further here. In that set-up, one creates which-way knowledge by tagging the system, which means a superposition was reversibly modified to become an entangled superposition between two degrees of freedom—for the Stern-Gerlach system this is done with an internal state of the atom. For the two-slit system, we do it with polarizers. If we add a vertical polarizer at slit A and a horizontal polarizer at slit B, then the state of the system after going through the slits is given by $\frac{1}{\sqrt{2}}(|\psi_A\rangle \otimes |V\rangle + |\psi_B\rangle \otimes |H\rangle)$. The creation of this entangled state is done by a measurement from the initial product state $|\psi\rangle \otimes |D\rangle$. Passing a photon through a polarizer is a measurement. It can be described in the standard Copenhagen paradigm, where the photon becomes entangled with the polarizer, and only the component with the polarization oriented along the polarizer passes through, leading to a projective measurement on the photon; this means we project with $|H\rangle\langle H|$ for horizontal polarization and $|V\rangle\langle V|$ for vertical polarization. Note that it has no classical pointer though, as is used in the standard description of a quantum measurement. If the incident photon enters with a diagonal polarization, the state of the photon changes via $|\psi\rangle \otimes |D\rangle \rightarrow \frac{1}{\sqrt{2}}(|\psi_A\rangle \otimes |V\rangle + |\psi_B\rangle \otimes |H\rangle)$ because $|H\rangle\langle H|D\rangle = \frac{1}{\sqrt{2}}|H\rangle$, and similar

for the vertical polarization. This is an interesting example for how to create an entangled state, because it does so via measurement from an initial product state. The irreversibility of the measurement is clear for two reasons—the amplitude is reduced to take into account the 50% of the photons absorbed by the polarizers, and one cannot reconstruct the original state from this state using any unitary operation. We also call this state a tagged state, but because it was constructed via a measurement, it is not a reversible tag that can be completely undone (on the other hand, in the Stern-Gerlach experiment, it is simple to create tags from unitary operations that can be completely undone). But, now when we measure at the screen, all interference effects are gone, because the polarization is projected with the operator $|V\rangle\langle V| + |H\rangle\langle H| = \hat{I}$, and the two components of the entangled superposition are orthogonal in the tensor-product because $\langle V|H\rangle = 0$. The probability then becomes $\frac{1}{2}(|\langle x|\psi_A\rangle|^2 + |\langle x|\psi_B\rangle|^2)$, which no longer has any interference effects. This shows an important general principle, that when we measure an entangled state, whether created by reversible tagging or by measurement, the presence of the tag in an entangled superposition at the time of the final measurement removes the interference effects.

Having this information under one's belt now allows one to get to the very bizarre quantum behavior. We are now ready to discuss the delayed-choice experiments, introduced by John Wheeler [11]. Again, we focus on the two-slit experiment, although these ideas can also be implemented with Stern-Gerlach analyzers and analyzer loops. The way we remove the tag involves placing a second polarizer in front of the detector on the screen. If we orient it vertically (A) or horizontally (B), we only see the light from slit A or slit B, respectively. But, if we orient it along the diagonal, the interference pattern is restored, because we have transformed the entangled state to a product state (at the cost of absorbing half of the photons again). In other words, we have $\frac{1}{\sqrt{2}}(|\psi_A\rangle \otimes |V\rangle + |\psi_B\rangle \otimes |H\rangle) \to \frac{1}{2}(|\psi_A\rangle + |\psi_B\rangle) \otimes |D\rangle$, after projecting onto $|D\rangle\langle D|$, which is a product state that will show interference effects, because it no longer has the which-way information of an entangled state. The use of the Dirac notation allows one to dispel any belief that the quantum system goes back in time (retrocausality) when a polarizer is placed before the detector. It is not needed, as the results just follow straightforwardly from the formal treatment with projection operators in the Copenhagen fashion. The key is the transformation from an entangled state to a product state. This product state loses its tagging of the which-way information and hence it cannot tell which path the photon went on, so we have quantum interference when we perform the final measurement at the screen.

This type of analysis also allows one to discuss the so-called *separation fallacy* [12]. This occurs when one conflates creating a superposition state, or an entangled tag, with creating a mixed state due to the measurement of an entangled state. For example, moving through an analyzer loop with both pass-through gates open, is represented by an atom in a superposition state. But, placing a pass-through detector at just one arm (which detects an atom that passes through that arm, but does not destroy or otherwise perturb the atom), then puts the system into a *mixed state*, which cannot be easily undone. Another example we already mentioned is the case with polarizers at the slits of a two-slit experiment, where a measurement is used to create a tag. For a solid understanding of quantum mechanics, one needs to be able to distinguish these events and understand when a tag is applied without measurement (because it is reversible and the norm of the state has not changed) versus by a measurement, which is not reversible, and often involves a change in the norm of the state). These distinctions are subtle, and often challenging to sort out without seeing a number of different examples that show the different possible behavior. For example, light incident on a beam splitter, is a great example to carefully distinguish superposition (or tagging) versus measurement.

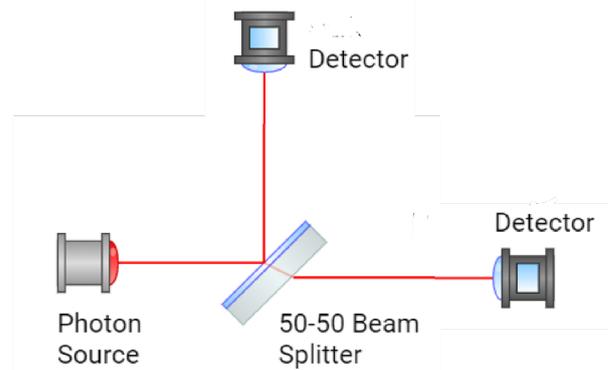

**Figure 3.** Schematic of a beam splitter to illustrate the difference between tagging, measurement, and the separation fallacy.

Figure 3 shows a schematic of a simple experiment with a beam splitter that illustrates the separation fallacy, which is a common misconception. A beam splitter has two input ports and two output ports—here, we show only one of the input ports—while each output port is sent to a single-photon detector. The beam splitter creates an indeterminate superposition of the photon along the paths corresponding to each output port. If it is a balanced beam splitter, the amplitude of the quantum state is the same for each output port. How do we know the beam splitter created a superposition, rather than a measurement? We know because the process is reversible. If we remove the detectors and bring in two additional mirrors and an additional beam splitter (with a flipped orientation) to create a Mach-Zehnder interferometer, we know that if the two paths through the interferometer are identical in length, then the photon emerges along only one output port of the interferometer (leading to the so-called bright detector). This is the same phenomenon as in the analyzer loop—in this case, the Mach-Zehnder interferometer does nothing—hence, going through the beam splitter was fully reversible, and it cannot be considered to be a measurement. (In teaching this concept, the student would need to have analyzed a Mach-Zehnder interferometer before having this discussion). But back to the original set up. The photon will transform to $|\psi\rangle \to \frac{1}{\sqrt{2}}(|up\rangle + |right\rangle)$ after the beam splitter. Note that this is a superposition, not a tagged entangled state (one could tag the state if one uses a polarizing beam splitter, but we do not do that here).

Let us go back and analyze the experiment. From a Copenhagen interpretation, one would say the photon has a well defined position as it moves toward the beam splitter, then it transforms to a superposition after the beam splitter, and the results of the measurement (which detector fires) will be random, but each detector will fire about half of the time. The state of the photon, including what path it is on is unknown until it has been measured—this is where the collapse of the wavefunction occurs. But this is just one interpretation. Another one that could be used is to say if the photon triggered the ``up'' detector, then we know the photon must have traveled on the up path to get there (how else could it go), so we can infer a well-defined path *a posteriori*. This is the type of analysis that would be used in a consistent history interpretation [13] of the experiment and is also what is usually invoked in a watched two-slit experiment with detectors at the slits—here, the detectors are moved away from the beam splitter, but the logic is identical. So, which is right? We cannot tell. Because both interpretations give the same experimental results, and if we remove the detectors, the consistent history will change, so a consistent history interpretation of, say a Mach Zehnder interferometer, will be different from the interpretation of the beam splitter with trailing detectors. What the separation fallacy tells us is that we cannot say that if the detctors are in place, then the photon chooses one path or another, and if we remove the detectors before the photon gets there, then we have a retrocausal change of behavior and the system changes to moving in superposition. One might argue, but why try to teach such complications to teachers? The answer is because these are precisely the types of questions their students will ask, and teachers need to be empowered to be able to answer them.

We end this section with a discussion of measurement, because this is one of the most poorly described topics in quantum textbooks. The main issue is textbooks describe measurement in a von Neumann fashion [13] via the quantum object becomes entangled with an apparatus, the apparatus then is connected to a pointer which indicates the result, and often the environment is brought in to reduce fluctuations and produce a definite answer because the pointer is macroscopic. Then, the Copenhagen interpetation is employed with wavefunction collapse to describe the experimental measurement, and the state of the system post measurement. As a short summary, this is the collapse to an eigenstate of the operator being measured, with the result of the measurement being the eigenvalue associated with that eigenstate. Here is the problem—very few experiments can be described in this fashion. The polarizer experiment we described, may have entanglement of the photon with the polarizer to change the polarization as it moves through the device, but there is no pointer; we only know the polarization after the polarizer because we know how the polarizer is oriented.

In fact, most measurements don't work in the von Neumann paradigm at all. Most quantum experiments are *counting experiments*, of one form or another, which usually destroy the particle being counted. Consider a single-photon photomultiplier tube. It detects the photon by detroying it and creating an electron (via the photoelectric effect), whose signal is amplified via a cascade process and then measured. There is no entanglement, and no pointer. What about measuring frequencies, or spectra? A spectroscope measures the angle the photon diffracts from a diffraction grating (after detecting the diffracted photon, hence a counting experiment). Is there entanglement, a pointer, or dissipation with the environment? No, because it is only measured when we destroy the photon to detect it. The collapse picture also fails when considering a time-of-flight measurement. Here, we have a trigger that starts a clock, then we let the particle evolve until it is detected in a detector (which is usually a destructive particle counter). We record the time evolved during the flight and the distance from the source to the detector of the flight to determine the velocity, and hence the momentum of the particle. But at the instant of the measurement, I know the position of the particle (because it is in the detector) and I know its momentum, by inference (because I know the mass, the time, and the distance traveled). So, how does collapse and Copenhagen work for this? As far as I can see, it does not. How does uncertainty work? The uncertainty arises from the fact that if I repeat the experiment from the same initial state, I will measure a distribution of times, hence I will obtain a distribution of momenta, which determines the uncertainty in momenta for the *initial state* of the particle. These end up being great examples of real experiments, that can be discussed instead of the idealized (and often impractical) experiments that exhibit the von Neumann methodology in quantum textbooks. Students need to understand how *real* experiments work if they are to become quantum aware; especially if they will move into the field of quantum sensing. Their teachers need to know this too.

Things get even more perplexing when we discuss interaction-free measurements. Probably the simplest interaction-free measurement is the Millikan oil-drop experiment. There, we measure the charge of single electrons on atomized droplets of oil, suspended between two capacitor plates. There is no entanglement with a measurement apparatus, no collapse, no pointer (unless you view the voltage as a pointer). If we get more sophisticated, the quantum-seeing-in-the-dark experiment [15] allows one to see an object with photons that never interact with them! One can even distribute quantum keys using this mechanism [16]. These experiments are simple to explain once one has developed quantum intuition and understands the principles of superposition, entanglement, complementarity, tagging, and measurement. They require no high-level math. They can be developed to be used in teacher preparation. We need to focus on the modern experiments that illustrate the bizarre behaviors of individual quanta. For this is what will be exploited in the second quantum revolution and is what the quantum-aware individual needs to know.

It may seem like these ideas are too advanced and cannot be taught to novices, especially using only high-school algebra, but this is not the case, as shown in the references discussed above. There are even materials that cover these concepts at a level appropriate for middle school students (Terry Rudolph's *Q is for Quantum* [4] and Michael Raymer's *Quantum Physics: What Everyone Needs to Know* [17]). Presenting quantum material in this fashion requires one to think about what the material really means. This is where learning truly takes place and where teachers can build confidence that they are mastering the material. But, it does take some time to get there. It cannot be done too quickly.

## Strategies for teaching conceptual ideas and learning goals

The United States Government, led by the National Science Foundation, held a workshop that developed nine key concepts for future quantum information science learners [18]. The goals we have developed for the conceptual ideas of quantum mechanics are anchored in these nine concepts, but do not focus on all of them, nor are they formulated strictly in terms of those concepts. Instead, they are organized around the notion of understanding experiments, and the formal ideas needed to explain those experiments within a quantum-mechanical framework. The nine key concepts include the following: (i) defining what quantum information science is; (ii) defining a quantum state; (iii) explaining how we measure quantum systems; (iv) defining what a qubit is; (v) explaining how quantum systems may be entangled; (vi) how they are coherent; (vii) defining what quantum computing is; (viii) what quantum communication is; and (ix) what quantum sensing is.

Our five conceptual learning goals for teachers lean toward the quantum sensing side of quantum information science and are the following:
(1) Be able to describe how the principle of superposition underlies indeterminacy and use the quantum superposition of states to analyze properties of spin (Stern-Gerlach experiments) and light (two-slit experiment and the Mach-Zehnder interferometer).
(2) Distinguish between events, alternative ways an event occurs, tagging, and measurement; be able to use the mathematical formalism of quantum mechanics to describe these different phenomena and how they allow for delayed-choice experiments.
(3) Describe how entanglement requires quantum mechanics to violate local realism and how experiments verify that this occurs.
(4) Explain how an interaction-free experiment allows for quantum seeing in the dark and the details of how such an experiment is carried out.
(5) Describe how real measurements of single quanta are made. Distinguish between projective measurements that can be described by the Copenhagen interpretation and measurements that are not projective (and often are inferred experiments or counting experiments).

These learning goals map to the nine concepts as follows: (1) maps to (ii), (iii), and (iv); (2) maps to (iii), (iv) and (v); (3) maps to (ii) and (v); (4) maps to (iii) and (ix); and (5) maps to (iii), (v) and (ix).

It almost appears that these learning goals are not reachable, that one is trying to teach too much to the teachers. Do they really need this? Let me describe why the answer is assuredly "yes they do." Quantum mechanics ideas are often bizarre. Take, for example the notion of superposition and indeterminacy. In classical logical thinking if I have two options, it must be one or the other. Not so in the quantum world. Here I can have a quantum system not have a definite property of one or the other. It doesn't even need to be a 50-50 split, it can be varying probabilities for one or another to occur, but neither being definite until measured. This requires a shift in the way teachers think and the only way to achieve a shift in thinking is to immerse oneself into the bizarre behavior of the quantum world with deep enough study that the ideas will stick.

We have found that effective teaching of these ideas to a quantum illiterate, but well-educated public, requires one to use a wide range of different techniques. This includes lectures, computer-based animations and tutorials, readings, numerical problem solving, development of models for physical behavior, discussion of philosophical ideas related to the content, and writing essays that synthesize different ideas together. It can be done in an on-line setting over a 4-6 week period with 7-10 hours per week of effort.

It is critically important to develop the ideas of the second quantum revolution. Namely, describe what individual quanta are (electron, photon, atom, etc.) and how we can measure their properties even if we cannot see them. It is important to properly describe what experiments can be done on these systems concretely. Let us not abstractly describe an experiment that has never, and will never be possible to be done. Let us instead describe experiments that are commonly used to determine properties of single quanta and that might even be used to control the properties of quantum systems. This approach naturally leads into ideas related to quantum computing hardware and quantum communications hardware, which routinely test and verify these types of ideas and which require an understanding of the five key conceptual ideas highlighted in this work. Modern experiments include interaction-free experiments and experiments that use quantum methodologies to improve the accuracy (such as using squeezed states in interferometry). Teachers need to be strong enough in working with these ideas that they will feel comfortable working with students and feel comfortable answering the many questions their students will have.

Completing such a program of study requires a rethinking of how we teach quantum mechanics, This is sorely needed, because the current paradigm for teaching quantum mechanics was developed in the 1930s and solidified in the 1950s—it has hardly changed since then. The second quantum revolution requires a rethinking of how to introduce these ideas. We have to jettison differential equations, jettison statements such as that "quantum mechanics is too hard to explain," or that it can "only be understood by the math." We need *not* "shut up and calculate." We need to be provided the tools to allow us to *think and conceptualize*. It can be done and it must be done to be successful.

## Conclusions

This paper expresses some strong opinions. They are the opinion of the author, but they are based on many hundreds of hours of instruction with students of widely different backgrounds and from many different countries. We need to listen to what physics education research has taught us over the past three decades—that a focus on conceptual ideas enables students to work better with formal developments and with problem solving. It helps them think like a quantum mechanic. The second quantum revolution calls upon us to redesign how we teach quantum mechanics, especially if we want to introduce quantum ideas earlier in the curriculum. We have identified five key conceptual ideas (superposition, entanglement, complementarity, tagging, and measurement) that are grounded in the nine key concepts for quantum information science learners, and are focused on the novel ideas presented within the second quantum revolution. These concepts have been developed into five learning goals that illustrate, in increasing complexity, how the five conceptual ideas can be employed to understand quantum phenomena. We owe it to the teachers, who will be introducing many students to quantum mechanics for the first time, to instill within them a strong conceptual background in quantum ideas that will allow them to have confidence in teaching material that often feels foreign and confusing. In our opinion, this is the only way forward. Many ideas and many materials exist to carry this out, but many more need to be developed to be successful. We hope this paper will help motivate others to join the effort in developing the needed curriculum.

## Acknowledgments

Ideas discussed in this paper were developed primarily within the edX courses *Quantum Mechanics for Everyone* and *Quantum Mechanics*. We acknowledge Lucas Vieira and Dylan Cutler for developing the computer tutorials, whose still images are used in the paper.

## Funding

We acknowledge the support of the National Science Foundation under grant number PHY-1915130. We also acknowledge support from the McDevitt bequest at Georgetown University.

## Ethical Statement

The author declares that he has no conflicts of interest.